\documentclass[11pt,a4paper]{article}
\pdfoutput=1
\usepackage{jheppub}
\usepackage[utf8]{inputenc} 
\usepackage{multirow}    
\usepackage{xcolor}
\usepackage{graphicx}
\usepackage{amsmath}
\usepackage{amssymb}
\usepackage{braket} 
\usepackage{hyperref}
\usepackage{frontespizio}
\usepackage{comment}  
\usepackage{epsfig}
\usepackage{float}
\usepackage{bbold}
\usepackage[title]{appendix}
\usepackage{multirow} 
\usepackage{tikz}
\usetikzlibrary{decorations.pathmorphing}
\usepackage{varwidth}
\usepackage{tikz}
\usetikzlibrary{backgrounds,patterns,calc}
\usepackage{xparse}  
\usepackage{subfigure}  
\usepackage{bbm}   
\usepackage{mathabx}

\usepackage{jheppub} 

\title{Factorisation Homology of Class ${\cal S}$ Theories}

\author[]{Veronica Pasquarella}
\affiliation{Department of Applied Mathematics and Theoretical Physics (DAMTP)}
\affiliation{University of Cambridge, \\ Wilberforce Road, CB3 0WA, Cambridge, UK}

\emailAdd{vp360@damtp.cam.ac.uk}

\abstract{Based on recent advancements in algebraic geometry, algebraic topology, and higher-categorical structures, we show how ground state degeneracies in closed stratified manifolds can be used for describing class ${\cal S}$ theories whose AGT dual requires generalised Moore-Segal bordism operators lacking reparametrisation-invariance.}

\makeatletter
\gdef\@fpheader{}
\makeatother
\begin{document} 
\maketitle

\section{Introduction}

\medskip  

\medskip

This is the second of a pair of mathematically-oriented papers by the same author explaining the proposed correspondence in between the role played by magnetic quivers and Drinfeld centers, \cite{Pasquarella:2023exd}. In \cite{Pasquarella:2023ntw}, we showed that, it is possible to generalise the Moore-Segal construction of bordism operators to cases involving lack of reparametrisation invariance. As argued in \cite{Pasquarella:2023ntw}, this corresponds to dropping dualisability of the 2-categories involved, thereby enabling to account for systems describing two different class ${\cal S}$ theories separated by a non-invertible defect, \cite{Bashmakov:2022uek,Pasquarella:2023deo}\footnote{According to the Alday-Gaiotto-Tachikawa (AGT) correspondence, \cite{Gaiotto:2009we,Alday:2009aq} one can map the description of class ${\cal S}$ theories to 2D TFTs, namely functors with bordism 2-categories as source, and target hyperk$\ddot{\text{a}}$hler quotient as target.}.

In \cite{Pasquarella:2023exd}, we argued in favour of the following statements:  

\begin{itemize}  

\item Lack of reparametrisation-invariance of bordism operators signals the presence of intrinsic non-invertible defects separating different class ${\cal S}$ theories.

\item Defining the Drinfeld center for a system of composite class ${\cal S}$ theories separated by non-invertible defects constitutes a nontrivial generalisation of an S-duality statement.

\end{itemize}

Mostly relying upon \cite{Moore:2006dw}, we highlighted the importance of the construction of cobordism operators, emphasising their dependence on the conformal structure of the Riemann surface. We then turned to the discussion of a particular 2D TFT valued in a symmetric monoidal category, namely the maximal dimension Higgs branch of class ${\cal S}$ theories. After briefly reviewing the properties outlined in \cite{Moore:2011ee}, we proposed their generalisation for the case in which the target category of the $\eta_{_{G_{_{\mathbb{C}}}}}$-functor is a hyperk$\ddot{\text{a}}$hler quotient. We concluded outlining the possible extension of this treatment towards a mathematical formulation of magnetic quivers within the context of Coulomb branches of 3D ${\cal N}=4$ quiver gauge theories.

In this new work, we take a more algebro-geometric approach in explaining the mathematical structure underlying supersymmetric quiver gauge theories, with a particular focus on dualities and their mutual relations in terms of higher-categories.

The main focus of this paper is that of explaining the higher categorical structures that are needed for describing the invariants associated to specific supersymmetric quiver gauge theories admitting a magnetic quiver description.  

As will be explained in depth, the crucial role is played by factorisation homology, \cite{fh}. We will not attempt to provide an exhaustive definition here, and refer the reader to the related section in the core of this work, instead. What we can anticipate, though, is that the role of factorisation homology is that of assigning invariants to topological manifolds, and, when applied to the AGT correspondence, provides a tool for evaluating the 4-sphere partition function of class ${\cal S}$ theories. Specifically, as we shall see, factorisation homology allows us to relate categorical dualisability and full-extendibility\footnote{And is therefore in line with the connection between dualities and categorical structures.}, \cite{Lurie:2009keu}. Within the context of non-Lagrangian theories, such as 6D ${\cal N}=(2,0)$ SCFTs, \cite{Witten:1995zh,Witten:2007ct}, this is particularly relevant since it introduces the notion of ring objects, of which Hilbert series are one of the examples of most interest to us.

As an algebraic variety, we focus on the calculation of the Hilbert series on Coulomb and Higgs branches, with the latter being related by the geometric invariant theory (GIT) construction, \cite{Deligne-Mumford}. GIT being a recipe for realising 3D mirror symmetry, \cite{Intriligator:1996ex}, we show how it can be suitably extended for cases where mirror duals are not necessarily in a 1-to-1 correspondence. In relation to \cite{Pasquarella:2023exd}, this nicely fits with the magnetic quiver prescription for describing the Higgs branch of certain supersymmetric quiver gauge theories with non-primitive ideals\footnote{Namely the 2-cone structure, \cite{Cabrera:2019izd,Ferlito:2016grh}.}. 
Specifically, we show how the 2D description of disk algebras can be suitably adopted for describing Hitchin systems associated to joint class ${\cal S}$ theories, \cite{Neitzke:2014cja}. In doing so, we will be making use of the full-extendibility/gauging argument outlined in section \ref{sec:334}.




\begin{figure}[ht!]  
\begin{center}    
\includegraphics[scale=1]{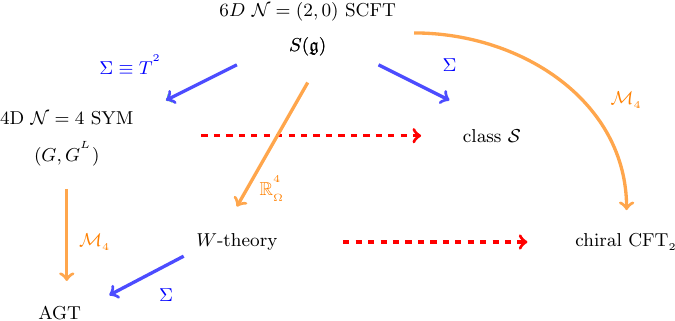} 
\caption{\small Reproduction of part of a diagram first presented in \cite{Moore}.}    
\label{fig:FH2}  
\end{center}  
\end{figure}

The present work is structured as follows: 

\begin{enumerate}

\item Section \ref{sec:DCFMQS} is devoted to explaining the role played by factorisation homology in describing the invariants of a given TFT. In particular, this prepares the stage for the following sections, where we will be dealing with the applicability of this formalism to the specific case of class ${\cal S}$ theories.

\item In section \ref{sec:334}, we then turn to explaining the relation between categorical dualisability and full-extendibility. In particular, this nicely relates to the notion of Drinfeld centers and gauging categorical structures.

\item Section \ref{sec:32} recapitulates the essential tools required for performing GIT quotient constructions, highlighting its importance for realising 3D mirror symmetry. In particular, we focus on the calculation of the Hilbert series as specific examples of algebraic varieties of crucial interest in the study of categorical dualities for supersymmetric quiver gauge theories.

\item The concluding section, namely \ref{sec:3}, shows how Coulomb and Higgs branches from class ${\cal S}$ Hitchin systems carry the same information as the gaugeable algebras leading to SymTFT constructions, discussed in \cite{Pasquarella:2023deo}. We conclude outlining interesting open questions and future directions of investigation by the author.

\end{enumerate}

\medskip    

\medskip

\section{Factorisation Homology in context}   \label{sec:DCFMQS}  

In this first subsection, which is mostly revision of well-known literature, \cite{tofh}, we provide a brief mathematical digression, needed for introducing factorisation homology and its utility for evaluating the index of a 2D TQFT. In doing so, we explain the notions of Hochschild cohomology, stratified spaces, and disk algebras. In the following sections we will see that, combining this formalism to algebro-geometric techniques, leads to interesting new findings.


\subsection*{Cohomology and Homotopy Equivalence}  

In homology theory and algebraic topology, cohomology is a general term for a sequence of abelian groups, usually one associated with a topological space, often defined from a cochain\footnote{Cochains are functions on the group of chains in the homology theory.} complex. Cohomology can be viewed as a way of assigning richer algebraic invariants to a space with respect to homology. 

Some versions of cohomology arise by dualising the homological construction. Some of their properties are formulated in a similar fashion with respect to their homological counterpart. However, cohomology enjoys a richer structure, which is lacking in homology, which we now turn to explain. 

For a given topological space $X$, and a commutative ring, $R$, the cup product is a bilinear map 

\begin{equation}  
H^{^i}(X,R)\ \times \ H^{^j}(X,R)\ \rightarrow\ H^{^{i+j}}(X,R).      
\end{equation}  

This product defines the cohomology ring of $X$ as follows

\begin{equation}  
H^{^{\bullet}}(X,R)\ \equiv\ \bigoplus_{_i} \ H^{^j}(X,R).       
\end{equation}  

It is graded commutative in the sense that, given $u\ \in\ H^{^i}(X,R)$, $v\ \in\ H^{^j}(X,R)$, 

\begin{equation} 
uv\ \equiv\ (-1)^{^{ij}}\ vu.  
\end{equation}

It follows that, if two spaces are homotopy equivalent, their cohomology rings are isomorphic.   

Homotopy equivalence can be expressed in terms of looping by representation spheres. In particular, a homotpy equivalence between topological spaces $X$ and $Y$ is a morphism $f:X\rightarrow Y$ which has a homotopy inverse, hence such that there exists a morphism $g: Y\ \rightarrow X$ and homotopies 

\begin{equation}  
g\ \circ\ f\ \sim\ \mathbf{1}_{_X}\ \ \ \text{and} \ \ \ f\ \circ\ g\ \sim\ \mathbf{1}_{_Y},   
\end{equation}
thereby making homotopy equivalence very similar to the notion of categorical equivalence.

For a given topological space, $X$, elements of $H^{^i}(X)$ can be thought of as represented by codimension-$i$ subspaces of $X$ that can move freely on $X$. For the case at hand, an element of $H^{^i}(X)$ is defined by a continuous map 

\begin{equation}  
f:\ X\ \rightarrow\ M,         
\end{equation}
with pullback 
\begin{equation}  
f^*:\ H^{{\bullet}}(M,R)\ \rightarrow\ H^{\bullet}(X,R),         
\end{equation}
and a closed codimension-$i$ submanifold $N$ of $M$ with an orientation in the normal bundle. As such, the class $f^*([N])\ \in H^{^i}(X)$ can be thought of to lie in the subspace $f^{^{-1}}(N)$ of $X$, since $f^*([N])$ restricts to $0$ in the cohomology of the open subset $X-f^{^{-1}}(N)$. The cohomology class $f^*([N])$ is free to move in $X$ in the sense that $N$ could be replaced by any continuous deformation of $N$ inside $M$.

\subsection{Hochschild cohomology and cochain complexes}   

Given a finite group action on a suitably enhanced triangulated category linear over a field, \cite{Perry}, Hochschild cohomology and group actions establish a formula for the Hocschild cohomology of the category of invariants, assuming the order of the group is coprime to the characteristic of the base field.

For a $k$-linear category with $G$-action,

\begin{equation} 
\underset{g\in G}{\bigoplus}\ \phi_{_g}:\ {\cal C}\ \rightarrow\ {\cal C} 
\end{equation}
induces the norm functor 

\begin{equation} 
\begin{aligned}
&N_{_m}: \ {\cal C}_{_G}\rightarrow{\cal C}^{G} \\
&\\
&\ \ \ \ \ \ \ \ x\ \mapsto\ \underset{g\in G}{\sum}\ g(x)       
\end{aligned}
\end{equation}
characterised by the existence of a factorisation property

\begin{equation}  
N_{_m}\circ q: \ {\cal C}\rightarrow {\cal C}^{^G}.  
\end{equation}    

Importantly, $N_{_m}$ is an equivalence.  

If ${\cal C, D}$ are $k$-linear categories, then the $k$-linear exact functors from ${\cal C}$ to ${\cal D}$ form the objects of the $k$-linear category Fun$_{_k}({\cal C},{\cal D})$. 

Let ${\cal C}$ be a $k$-linear category, and let $\phi:\ {\cal C}\rightarrow{\cal C}$ be an endofunctor. The Hochschild cochain complex of ${\cal C}$ with coefficients in $\phi$ is defined as 

\begin{equation}  
HC^{^{\bullet}}({\cal C},\phi)\ \overset{def.}{=}\ \text{Map}_{_{\text{Fun}_{_k}({\cal C,C})}}(\mathbb{1}_{_{\cal C}},\phi)\ \in\ \text{Vect}_{_k}.  
\end{equation}  

The Hochschild cohomology $HH^{^{\bullet}}({\cal C},\phi)$ of ${\cal C}$ with coefficients in $\phi$ is the cohomology of this complex. For the case in which $\phi\equiv\mathbb{1}_{_{\cal C}}$, 

\begin{equation}   
HC^{^{\bullet}}({\cal C})\ \overset{def.}{=}\ HC^{^{\bullet}}({\cal C},\mathbb{1}_{_{\cal C}}) 
\ \ \ ,\ \ \ 
HH^{^{\bullet}}({\cal C})\ \overset{def.}{=}\ HH^{^{\bullet}}({\cal C},\mathbb{1}_{_{\cal C}}) 
\end{equation}   
define the Hochschild cochain complex and the Hochschild cohomology of ${\cal C}$, resepctively. The latter comes equipped with a natural algebra structure, induced by composition in $\text{Map}_{_{\text{Fun}_{_k}({\cal C,C})}}(\mathbb{1}_{_{\cal C}},\mathbb{1}_{_{\cal C}})$.

Note that, given a $k$-linear category $\cal C$ with an action by a finite group $G$, there is an induced action of $G\times G$ on $\text{Fun}_{_k}({\cal C,C})$ consisting in the following. An element $(g_{_1},g_{_2})\ \in\ G\times G$ acts on $\text{Fun}_{_k}({\cal C,C})$ by sending $F:\ {\cal C}\rightarrow {\cal C}\ \in\ \text{Fun}_{_k}({\cal C,C})$ to $\phi_{_{g_{_2}}}\circ F\circ \phi^{^{-1}}_{_{g_{_1}}}$. Via the diagonal embedding $G\ \subset\ G\times G$, this restricts to the conjugation action of $G$ on $\text{Fun}_{_k}({\cal C,C})$.   

A main theorem in \cite{Perry} states that, given $\cal C$ a $k$-linear category with an action by a finite group $G$, and assuming the order of $G$ is coprime to the characteristic of $K$, then there is an isomorphism 

\begin{equation}   
HH^{^{\bullet}}({\cal C}^{^G})\ \simeq\ \left(\underset{g\in G}{\bigoplus}\ HH^{^{\bullet}}({\cal C},\phi_{_g})\right)^{^G}, 
\end{equation}
with $\phi_{_g}: {\cal C}\rightarrow {\cal C}$ an autoequivalence corresponding to $g\in G$, and the $G$-action on the right side induced by the conjugation action of $G$ on $\text{Fun}_{_k}({\cal C,C})$. In particular, 

\begin{equation}   
HH^{^{\bullet}}({\cal C})^{^G}\ \hookrightarrow\ HH^{^{\bullet}}({\cal C}^{^G}).
\end{equation}

We will come back to this in section \ref{sec:32} when discussing the GIT quotient and the Kirwan map.

\subsection{Factorisation homology of stratified spaces} \label{sec:fhss}


Factorisation homology satisfies a defining property called $\otimes$-excision, which determines $\int_{_M}{\cal A}$ up to equivalence, \cite{tofh},

\begin{equation} 
\boxed{\ \ \ \int_{_M}{\cal A}\ \simeq\ (H,u_{_{\Sigma}})  \color{white}\bigg]\  \ },   
\end{equation}
with such property being a special case of the pushforward property. It allows to reduce the computation of the factorisation homology of a surface to that of a lower-dimensional manifold, and, eventually, to that of a 0-dimensional manifold. 

The only known global observable on $\Sigma$ for an anomaly-free TO is the ground state degeneracy. The theory of factorisation homology gives a powerful tool for calculating it. When a closed stratified surface is decorated by anomaly-free topological defects of codimension 0,1,2, the theory of factorisation homology gives us a powerful tool for calculating the degeneracy of the ground state. As an important bi-product, FH provides a powerful tool for calculating global observables on surfaces for anomalous TOs. Indeed, relation between FH and TOs enables to make direct contact with the theoretical physics setup of interest to us.

There is a well-defined fully faithful functor 

\begin{equation}  
\mathfrak{Z}:\ \text{UMTC}^{^{\text{ind}}}\ \rightarrow\ \text{UMTC}  
\label{eq:drinf}
\end{equation} 
acting on objects and morphisms as follows

\begin{equation} 
{\cal C}\ \rightarrow\ \mathfrak{Z}({\cal C})\ \ \ ,\ \ \ _{_{{\cal C}}}{\cal M}_{_{{\cal D}}}\ \mapsto\ \mathfrak{Z}({\cal M})\ \overset{def.}{=}\ \text{Fun}_{_{{\cal C}|{\cal D}}}({\cal M},{\cal M}).     
\end{equation}  

For ${\cal C}$ a UMTC, \eqref{eq:drinf} defines the Drinfeld center of ${\cal C}$, $\mathfrak{Z}({\cal C})$, a canonical braided monoidal category such that 

\begin{equation}   
\mathfrak{Z}({\cal C})\ \simeq\ {\cal C}\ \boxtimes\ \overline{\cal C}.    
\end{equation}

More explicitly, Th.2.5 in \cite{tofh} implies that, given ${\cal C}, {\cal D}, {\cal E}$ UMFCs, ${\cal M}$ a multi-fusion ${\cal C}-{\cal D}$-bimodule, and ${\cal N}$ a multi-fusion ${\cal D}-{\cal E}$-bimodule, the assignment 

\begin{equation}  
f\boxtimes_{_{{\mathfrak{Z}({\cal D})}}}g\ \mapsto\ f\boxtimes_{_{\cal D}}g    
\end{equation}   
defines an equivalence between two multi-fusion $\mathfrak{Z}({\cal C})-\mathfrak{Z}({\cal E})$-bimodules:  

\begin{equation}  
\text{Fun}_{_{{\cal C|D}}}({\cal M},{\cal M})\boxtimes_{_{{\mathfrak{Z}({\cal D})}}}\text{Fun}_{_{{\cal D|E}}}({\cal N},{\cal N})\ \simeq\ \text{Fun}_{_{{\cal C|E}}}({\cal M}\boxtimes_{_{{\cal D}}}{\cal N},{\cal M}\boxtimes_{_{{\cal D}}}{\cal N})  
\end{equation}
from which the following two corollaries follow:

\begin{equation}  
{\cal C}\boxtimes_{_{{\mathfrak{Z}({\cal C})}}}{\cal D}^{^{rev}}\ \simeq\ \text{Fun}_{_{H}}({\cal M},{\cal M}) 
\end{equation}  

\begin{equation}  
\boxtimes_{_{({\cal C}_{_{0}},...,{\cal C}_{_{n-1}})}}^{^{\circlearrowright}}({\cal M}_{_{1}},...,{\cal M}_{_{n}})\ \simeq\ \text{Fun}_{_{H}}(\cal P,P), 
\end{equation}  
with ${\cal P}$ a unique unitary category such that this correspondence holds.

Next, consider Mfld$_{_{n}}^{^{or}}$ to be the topological category whose objects are oriented n-manifolds whose objects are oriented n-manifolds without boundary. For any two oriented manifolds $M,N$, the morphism Hom$_{_{\text{Mfld}_{_{n}}^{^{or}}}}(M,N)$ is the space of all orientation-preserving embeddings $e: M\rightarrow N$ endowed with the compact-open topology. 

Denoting with ${\cal M}\text{fld}_{_{n}}^{^{or}}$ the symmetric monoidal $\infty$-category associated to the topological category Mfld$_{_{n}}^{^{or}}$, the symmetric monoidal $\infty$-category Disk$_{_{n}}^{^{or}}$, is the full subcategory of ${\cal M}\text{fld}_{_{n}}^{^{or}}$ whose objects are disjoint union of finitely-many n-dimensional Euclidean spaces equipped with standard orientation. 

If ${\cal V}$ is a symmetric monoidal $\infty$-category, $M$ an oriented manifold, and 

\begin{equation}  
\boxed{\ \ \ {\cal A}:\ \text{Disk}_{_{n}}^{^{or}}\ \rightarrow\ {\cal V} \color{white}\bigg]\ \ }   
\end{equation}  
an n-disk algebra in ${\cal V}$, the factorisation homology of $M$ with coefficients in ${\cal A}$ is an object of ${\cal V}$ given by  

\begin{equation}  
\int_{_{M}}\ {\cal A}\ \overset{def.}{=}\ \text{Colim.}\left(\left(\text{Disk}_{_{n}}^{^{or}}\right)_{_{/M}}\ \rightarrow\ \text{Disk}_{_{n}}^{^{or}}\ \xrightarrow{\cal A}\ {\cal A}\right),  
\end{equation}   
with $\left(\text{Disk}_{_{n}}^{^{or}}\right)_{_{/M}}$ denoting the category of $n$-disks embedded in $M$. If $\mathbb{1}_{_{n}}$ is the trivial $n$-disk albegra assigning to each $\mathbb{R}^{^n}$ $\mathbb{1}_{_{{\cal V}}}\ \in\ {\cal V}$, it follows that 

\begin{equation}  
\int_{_{M}}\mathbb{1}_{_{n}}\ \simeq\ \mathbb{1}_{_{{\cal V}}}.
\end{equation}

The canonical morphism $\mathbb{1}_{_{n}}\rightarrow{\cal A}$ then induces a morphism $\mathbb{1}_{_{{\cal V}}}\rightarrow\int_{_{M}}{\cal A}$, implying the factorisation homology $\int_{_{M}}{\cal A}$ is not just an object of ${\cal V}$, but also comes equipped with a 0-disk algebra structure, the latter playing a crucial role for ground state degeneracy calculations. 

Explicit calculation of factorisation homology as a 0-disk algebra is very useful. Indeed, factorisation homology on a closed stratified surface $\Sigma$ with anomaly-free coefficient ${\cal A}$, $(H,u_{_{\Sigma}})$, exactly gives the ground state degeneracy. of the same surface decorated by anomaly-free topological defects of codimension 0, 1, and 2 that are associated to ${\cal A}$, namely $u_{_{\Sigma}}$.  

\subsection*{Examples of stratified factorisation homology}

Given a stratified 2-disk, $M_{_1}$, with ${\cal A}$ a 2-disk algebra assigned to 2-cells, 1-disk algebras associated to 1-cells, and ${\cal P}$ as the unique 0-algebra assigned to the unique 0-cell, examples of its factorisation homology read as follows

\begin{equation}   
\int_{_{M_{_1}}}{\cal A}_{_1}\ \simeq\ {\cal A}\ \ \ ,\ \ \ \int_{_{M_{_1}\diagdown \{0\}}}{\cal A}_{_1}\ \simeq\ HH_{_{\bullet}}({\cal A})\ \overset{def.}{\equiv}\ {\cal A}\otimes_{_{{\cal A}\otimes{\cal A}^{^{op}}}}{\cal A},
\end{equation}
with $HH_{_{\bullet}}({\cal A})$ denoting the Hochschild cohomology.   

For anomaly-free coefficient systems, 

\begin{equation}   
\boxtimes_{_{({\cal C}_{_{0}},...,{\cal C}_{_{n-1}})}}^{^{\circlearrowright}}({\cal M}_{_{1}},...,{\cal M}_{_{n}})\ \simeq\ \text{Fun}_{_{H}}(\cal P,P).
\end{equation}

The unique 1-cell labelled by a closed multi-fusion ${\cal D}-{\cal C}$-bimodule ${\cal M}$. The two 0-cells are labelled by $({\cal D}, d)$ and $({\cal M},m)$, for $d\in{\cal D}$, and $m\in{\cal M}$, respectively. The closed multi-fusion ${\cal C-D}$-bimodule structure on ${\cal M}$ induces two monoidal functors 
\begin{equation}  
{\cal C}\ \xrightarrow{L}\ {\cal M}\ \xleftarrow{R}\ {\cal D}.  
\end{equation}  

\begin{figure}[ht!]  
\begin{center}    
\includegraphics[scale=1]{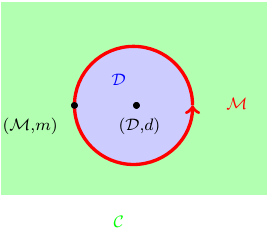} 
\caption{\small A stratified 2-disk with 2-cells labelled by UMTCs ${\cal C}$ and ${\cal D}$. The unique 1-cell is labelled by a closed multi-fusion ${\cal C}-{\cal D}$-bimodule, ${\cal M}$. The two 2-cells are labelled by $({\cal D},d), ({\cal M},m)$, for $d\in{\cal D}$ and $m\in{\cal M}$, respectively.}    
\label{fig:FH}  
\end{center}  
\end{figure}  

Given $(\mathbb{R}^{^2};\Gamma_{_0})$ a stratified 2-disk, the FH reads as follows 

\begin{equation}  
\int_{_{(\mathbb{R}^{^2};\Gamma_{_0})}}({\cal C},{\cal D};{\cal M};({\cal D},d),({\cal M},m))\ \simeq\ ({\cal C},L^{^{\text{V}}}(m\otimes R(d))),  
\label{eq:FH1}  
\end{equation}  
with $L$ the right-adjoint functor of $L$. If ${\cal D}\equiv{\cal C}^{^0}_{_{{\cal A}}}, {\cal M}\equiv{\cal C}_{_{{\cal A}}}$ for a connected commutative separable algebra ${\cal A}$ in ${\cal C}$, $d\equiv\mathbb{1}_{_{{\cal D}}}$, and $m\equiv\mathbb{1}_{_{\cal M}}$, then \eqref{eq:FH1} reduces to   

\begin{equation}  
\int_{_{(\mathbb{R}^{^2};\Gamma_{_0})}}\left({\cal C},{\cal C}^{^0}_{_{{\cal A}}};{\cal C}_{_{{\cal A}}}^{^0};\left({\cal C}_{_{{\cal A}}}^{^0},d\right),\left({\cal C}_{_{{\cal A}}},\mathbb{1}_{_{{\cal C}^{^0}_{_{{\cal A}}}}}\right)\right) \ \simeq\ ({\cal C},{\cal A}).
\label{eq:FH}  
\end{equation}


\subsection{Non-contractible loops} \label{sec:2.3}

In presence of non-contractible loops, 

\begin{equation} 
\int_{_{(S^{^2};S^{^1})}}({\cal C},{\cal D};{\cal K};\emptyset)\ \simeq\ (H, \text{Hom}_{_{{\cal C}\boxtimes{\cal D}}}(\mathbb{1}_{_{{\cal C}}\boxtimes_{_{\cal D}}} {\cal A})  
\end{equation}
with

\begin{equation}   
{\cal A}\ \overset{def.}{=}\ (L\ \boxtimes\ R)^{^V} (\mathbb{1}_{_{\cal K}}) 
\end{equation}
a commutative separable algebra in $\bar {\cal C}\ \boxtimes\ {\cal D}$. If ${\cal K}$ is not fusion, then 

\begin{equation} 
\text{dim Hom}_{_{{\cal C}\boxtimes{\cal D}}}(\mathbb{1}_{_{{\cal C}}}\boxtimes \mathbb{1}_{_{{\cal D}}}, {\cal A})\ =\ \text{dim Hom}_{_{k}}(\mathbb{1}_{_{{\cal K}}},  \mathbb{1}_{_{\cal K}})\ >\ 1. 
\end{equation}

When this happens, the corresponding topological order is not stable, and triggers an RG-flow.  

Via factorisation homology, one defines the $W$-matrix, $W^{^{\cal M}}$, for ${\cal M}$ as follows  

\begin{equation}  
(L_{_{\cal M}}\ \boxtimes\ R_{_{\cal M}})^{^V}(\mathbb{1}_{_{{\cal M}}})\ \simeq\ \underset{i\ \in\ O({\cal C}), \ j\ \in\ O({\cal D})}{\bigoplus}\ W^{^{\cal M}}_{_{ij}}\ \ i\ \boxtimes\ j^{^*}  
\end{equation}

\begin{equation}  
(L_{_{\cal M}}\ \boxtimes\ R_{_{\cal M}})^{^V}(\mathbb{1}_{_{{\cal M}\boxtimes_{_{{\cal D}}}{\cal N}}})\ \simeq\ \underset{i\ \in\ O({\cal C}), \ j\ \in\ O({\cal E})}{\bigoplus}\ (W^{^{\cal M}}W^{^{\cal N}})_{_{ij}}\ \ i\ \boxtimes\ j^{^*}, 
\end{equation}  
given $(\mathbb{R}\times S^{^1}; S^{^1}\ \cup\ ...\ \cup\ S^{^1})$ a stratified open cylinder with 2-cells labelled by UMTCs ${\cal C}_{_{0}}, ..., {\cal C}_{_{n}}$, and 1-cells (namely non-contractible loops) labelled by closed multi-fusion ${\cal C}_{_{i-1}}-{\cal C}_{_i}$-bimodules ${\cal M}_{_i}$, $i=1,...,n$, and no 0-cells.

Crucially, the above leads the definition of the ground state degeneracy, namely the index we are looking for

\begin{equation}  
\boxed{\ \ \ \text{dim}\  u \ \overset{def.}{=}\ \text{Tr}\ \left(W^{^{\cal M}}_{_1}W^{^{\cal M}}_{_2}...W^{^{\cal M}}_{_n}\right) \color{white}\bigg] \ }.    
\end{equation}   

From the AGT correspondence, we know that this is equal to the partition function of a class ${\cal S}$ theory. 

As we will see in the upcoming sections, we can map the issue of gluing different class ${\cal S}$ theories separated by a non-invertible defect, \cite{Bashmakov:2022uek,Pasquarella:2023deo}, to that of defining the ground state degeneracy of a 2D TO in presence of non-contractible loops.

\section{Dualisability and full-extendibility}   \label{sec:334}  

We now turn to build connection with two of our previous works, namely \cite{Pasquarella:2023deo, Pasquarella:2023ntw}. Specifically, we overview the relation between categorical dualisability and full-extendibility of a 2D TQFT. As we shall see, when applying this to  a specific algebraic variety, namely the Hilbert series, homological mirror symmetry, in its categorical formulation, results as a biproduct of full-extendibility. 

\subsection{Extending TQFTs}

In \cite{Lurie:2009keu}, Lurie proved that a TQFT is fully determined by what it assigns to a point. Furthermore, fully extended TQFTs with values in a symmetric monoidal $(\infty, n)$-category, ${\cal C}$, are equivalent to fully dualisable objects in ${\cal C}$. 

For a 2D TQFT, there are two main cases, \cite{lgmtap}:

\begin{enumerate}

\item   The TQFT is fully-extended, in which case the 2-functor reads as follows:

\begin{equation}  
{\cal Z}:\ \text{Bord}_{2,1,0}^{^{\sigma}}\ \rightarrow\ {\cal B},  
\end{equation}
where $\text{Bord}_{2,1,0}^{^{\sigma}}$ denotes the bicategories of points, 1-manifolds with boundaries, and 2-manifolds with corners. The target, ${\cal B}$, is most commonly associated with the bicategory of finite-dimensional $\textbf{k}$-algebras, bimodules, and bimodule maps, Alg$_{_{\textbf{k}}}$. From the cobordism hypothesis, it follows that the extended framed TQFTs with values in Alg$_{_{\textbf{k}}}$are classified by finite-dimensioanl separable $\textbf{k}$-algebras.

\item  The TQFT is not fully-extended, instead, if

\begin{equation}  
{\cal Z}:\ \text{Bord}_{2,1}^{^{\text{or}}}\ \rightarrow\ {\cal V},  
\end{equation}
with ${\cal V}$ a symmetric monoidal 1-category.

\end{enumerate}

A non-extended 2D TQFT can be extended to the point if there is a symmetric monoidal bicategory, $\cal B$, and an extended TQFT such that

\begin{equation}  
{\cal V}\ \simeq\ \text{End}_{_{\cal B}}\left(\mathbf{1}_{_{\cal B}}\right)\ \ \ \ \text{and} \ \ \ \ {\cal Z}_{_{ne}}\ \simeq\ {\cal Z}\bigg|_{{\text{End}_{_{\text{Bord}_{_{2,1,0}}^{^{\sigma}}(\emptyset)}}}},   
\label{eq:prescr}
\end{equation}
where $\mathbf{1}_{_{\cal B}}\ \in\ {\cal B}$ and $\emptyset\ \equiv\ \mathbf{1}_{_{\text{Bord}_{_{2,1,0}}^{^{\sigma}}(\emptyset)}}$.  

The prescription \eqref{eq:prescr} is clearly not unique. In particular, it depends on the target ${\cal B}$. 

If a non-extended 2D TQFT is a restriction of an appropriate defect TQFT ${\cal Z}_{_{ne}}^{^{\text{def.}}}$, then the non-extended TQFT can be extended to the point as long as we associate it with the bicategory ${\cal B}_{_{{\cal Z}_{_{ne}}^{^{\text{def.}}}}}$.

Note that this closely resembles the gauging prescription of SymTFTs, \cite{Freed:2022qnc}.


In building the understanding for how the setup of subsection \ref{sec:2.3} is of use for our previous works, \cite{Pasquarella:2023exd, Pasquarella:2023ntw, Pasquarella:2023deo}, we first need a brief mathematical digression, presented in the remainder of this intermediate section. We will then combine these tools with those of section \ref{sec:DCFMQS} in the concluding part of this work.

\subsection{Dualisability in categorical structures}  

A dualisable object in a symmetric monoidal $(\infty,n)$-category, ${\cal C}$, is fully dualisable if the structure maps of the duality unit and counit each admit adjoints, as well as higher-morphisms between them, up to level $(n-1)$, \cite{Lurie:2009keu}. 

By the Cobordism Hypothesis Theorem, \cite{Lurie:2009keu}, symmetric monoidal $(\infty, n)$-functors out of the $(\infty, n)$-category of cobordisms are characterised by their value at the point, which is a fully-dualisable object.

The notion of dualisability and full dualisability can be extended to the category of chain complexes, where an object is said to be dualisable if and only if it is a bounded chain complex of dualisable modules.

\subsection{The case of the Hilbert series} 

We now turn to the higher-categorical structure most relevant for our treatment, whose dualisability is the core topic of the present and following section.

In \cite{Perry}, the author explores many applications of representation theory of categories to sequences of groups arising in topology, algebra, and combinatorics. 

Given $(V_{_n})_{_{n\in\mathbb{N}}}$ a sequence of representations of groups $G_{_n}$ over a field $k$, it can be lifted to a category ${\cal C}$ whose objects (or isomorphism classes) are indexed by natural numbers, such that the automorphism group of the $n^{th}$ object of ${\cal C}$ is $G_{_n}$. A ${\cal C}$-representation, or ${\cal C}$-module, is a functor ${\cal C}\rightarrow$ Mod$k$. One attempts to use the representation theory of ${\cal C}$ to show that the whole sequence of representations is determined by a finite amount of data, and to discover universal patterns which the sequence must satisfy.

These patterns are often expressed in terms of generating functions. For example, one invariant of a sequence of representations is its sequence of dimensions, (dim $V_{_n})_{_n\in\ \mathbb{N}}$. Such data can be recorded as a generating function, called the Hilbert series of $(V_{_n})_{_{n\in\mathbb{N}}}$

\begin{equation}   
\text{HS}_{_V}(t)\ \overset{def.}{=}\ \underset{n}{\sum}\ \text{dim}_{_k}V_{_n}t^{^n}\ \in\ \mathbb{Z}[[t]].  
\end{equation}

In most cases, theorems from the representation theory of ${\cal C}$ imply that HS$_{_V}(t)$ is rational, with denominator of a certain type. For example, considering the category of finite dimensional vector spaces over $\mathbb{F}_{_q}$ and linear injections between them, if $V$ is a finitely generated $VI_{_q}$ module, and char $k \neq$ char $\mathbb{F}_{_q}$, then HS$(t)$ is rational, with denominator 

\begin{equation} 
w_{_d}(t)\ \overset{def.}{=}\ \overset{d-1}{\underset{j=0}{\prod}}\ \left(1-q^{^j}t\right).  
\label{eq:wp}   
\end{equation}  

Each factor in \eqref{eq:wp} is a Whitney polynomial of the poset of subsets of a $d$-element set.

Using methods from poset topology, construct a chain complex of ${\cal C}$-modules $k_{_d}(V), \forall\ d\in\mathbb{N}$, such that exactness of $k_{_d}(V)$ categorifies the equation

\begin{equation} 
w_{_d}(t)\ HS_{_V}(t)\ \equiv\ p(t),    
\end{equation} 
for $p(t)$ a polynomial. If $V$ is finitely generated, either $k_{_d}$ or a power of it applied to $V$ is exact. In particular, categorification explains why the denominator \eqref{eq:wp} appears.

\subsection*{Poset homology} 

Poset homology are chain complexes classifying rationality.
Extended literature on the subject explains there is a method for assigning a chain complex to a module over a combinatorial category, such that, if the complex is exact, the module has a rational Hilbert series. Furthermore, these complexes are known to satisfy Homology-vanishing theorems. Hasse diagrams, discussed in \cite{Pasquarella:2023exd}, are an example of chain complexes associated to poset representations. Homological mirror symmetry as an equivalence in between Hilber series on two categories, and rationality, are a direct biproduct of the defining features of poset homology. For the sake of completeness, we briefly overview the main statement of homological mirror symmetry, as needed for

\subsection*{Homological mirror symmetry} 

Homological mirror symmetry constitutes a particular example of dualisable categorical structures, \cite{Kontsevich:1994dn}. It consists in a proposed agreement between two categories, namely:

\begin{itemize}  

\item  Fukaya's $A_{_{\infty}}$-category, ${\cal F}(X)$, on the symplectic side,

\item and the derived category with Yoneda structure, $D^{^b}\mathfrak{Coh}(X^{^{\text{V}}})$, on the complex side. 

\end{itemize} 

For a given symplectic manifold, $X$, of even dimension equipped with a non-degenerate symplectic form $\omega\ \in\ \Omega^{^2}_{_{cl}}(X)$, a Fukaya category of a symplectic manifold, $X$, is an $A_{_{\infty}}$-category with Lagrangian submanifolds $L\ \subset\ X$ as objects, whose intersections define the Hom-space. 
An $A_{_{\infty}}$-category is a category with associativity condition ($A)$ relaxed without bound on degrees of homotopies ($\infty$). They are linear categories, i.e. their Hom-objects are chain complexes. A Lagrangian submanifold of a symplectic manifold, $L\ \subset\ X$, is a submanifold which is a maximal isotropic submanifold on which $\omega=0$. They constitute the leaves of real polarisations, and are, therefore, crucial elements of symplectic geometry.

There is an equivalence of categories between the category of $\mathbb{C}$-linear representations of a quiver ${\cal Q}$ and the category of left $\mathbb{C}{\cal Q}$-modules.

\begin{figure}[ht!]  
\begin{center} 
\includegraphics[scale=1]{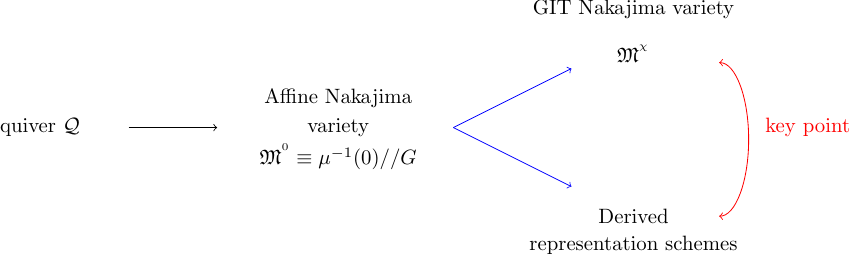} 
\caption{\small The upper and lower blue arrows correspond to geometric and algebraic resolutions of ${\mathfrak{M}^{0}}$, respectively. }  
\label{fig:GIT}  
\end{center}    
\end{figure}

  \subsection*{Moment map and higher homologies}

In general, homologies of derived representation schemes can be highly nontrivial. However, in this particular case, one can identify a necessary and sufficient condition for the vanishing of the higher homologies based on the flatness of the moment map, $\mu$, \cite{DAlesio:2021hlp}. In particular, in such reference, it was shown that the derived representation scheme DRep$_{_{v,w}}({\cal A})$ has vanishing higher homologies if and only if $\mu^{^{-1}}(0)\ \subset\ M({\cal Q},v,w)$ is a complete intersection, which happens only if the moment map is flat, \cite{mommap}. As explained in \cite{Pasquarella:2023exd,Pasquarella:2023ntw}, the requirement of the algebraic variety to be a complete intersection is crucial for the purpose of our treatment. In particular, it ensures the emergence of a 2-categorical structure, whose importance will be the core topic of section \ref{sec:3}.

\section{The GIT quotient and its deformation}  \label{sec:32}

Prior to delving in the explicit description of the duality structure arising from class ${\cal S}$ theories, to which section \ref{sec:3} is devoted, we will first overview some key preliminary algebro-geometric tools, specifically the definition of Coulomb and Higgs branches from geometric invariant theory, \cite{Nakajima:2022sbi,Gonzalez:2023jur,Braverman:2016pwk,Braverman:2016wma,Teleman:2022oiu,DP2021,Berest,Teleman:2018wac,GIT,Braverman:2017ofm}.

\subsection{Equivariant cohomology}

Equivariant cohomology, \cite{BFM}, also known as Borel cohomology, is a cohomology theory from algebraic topology which applies to topological spaces with a group action. It can be viewed as a generalisation of group cohomology. More explicitly, the equivariant cohomology ring of a space $X$ with action of a topological group $G$ is defined as the ordinary cohomology ring with coefficient ring $\Lambda$ of the homotopy quotient EG$\times_{_G}$X  

\begin{equation}  
H^{^{\bullet}}_{_G}(X;\Lambda)\ =\ H^{^{\bullet}}\left(EG\times_{_G} X;\Lambda\right).    
\end{equation}  

\subsection*{The Kirwan map}  

The Kirwan map, first introduced in \cite{FK}, states 

\begin{equation}  
H^{^{\bullet}}_{_G}(M)\ \rightarrow\ H^{^{\bullet}}\left(M//_{_p} G\right), 
\end{equation}  
where $M$ is a Hamiltonian $G$-space, i.e. a symplectic manifold acted on by a Lie group $G$ with a moment map 

\begin{equation}     
\mu:\ M\ \rightarrow\ \mathfrak{g}^{*}.  
\end{equation}   

$H^{^{\bullet}}(M)$ is an equivariant cohomology ring of $M$, i.e. the cohomology ring of the homotopy quotient $EG\times_{_G}M$ of $M$ by $G$.

\begin{equation}  
M//_{_p}G\ =\ \mu^{^{-1}}(p)/G  
\end{equation}   
is the symplectic quotient of $M$ by $G$ at a regular central value $p\in Z(\mathfrak{g}^{*})$ of $\mu$. It is defined as the map of equivariant cohomology induced by the inclusion 

\begin{equation}    
\mu^{^{-1}}(p)\ \hookrightarrow\ M   
\end{equation}
followed by the canonical isomorphism  

\begin{equation}  
H^{^{\bullet}}_{_G}\left(\mu^{^{-1}}(p)\right)\ =\ H^{^{\bullet}}\left(M//_{_p} G\right).
\end{equation}

\subsection{GIT}  

Geometric invariant theory (GIT), \cite{Deligne-Mumford} is a recipe for constructing quotients in algebraic geometry. An affine GIT quotient, or affine geometric invariant theory quotient of an affine scheme $X=\text{Spec} {\cal A}$ with an action by a group scheme $G$, is the affine scheme Spec$({\cal A}^{^G})$, namely the prime spectrum of the ring of invariants of ${\cal A}$, and is denoted by $X//G$.

The input of the GIT construction consists of:

\begin{enumerate} 

\item A complex reductive group, $G$.

\item A representation of $G$  

\begin{equation} 
X \ =\ \mathbf{A}^{^n}.  
\end{equation}

\item A character

\begin{equation}  
\theta:\ G\ \rightarrow\ G.   
\end{equation}

\end{enumerate}  

GIT takes this data and gives an open subset of $X$, namely the semistable locus  

\begin{equation}    
X_{_{\theta}}^{^{ss}}(G)\ \overset{\text{open}}{\subset}\ X, 
\end{equation}
from which one can define the stack quotient as follows 

\begin{equation}    
X//_{_{\theta}}(G)\ \overset{def.}{\equiv}\ \left[X_{_{\theta}}^{^{ss}}(G)\ /\ G\right]. 
\label{eq:ssq}
\end{equation}  

As \eqref{eq:ssq} suggests, a GIT quotient is therefore a categorical quotient of the locus of semistable points, i.e. the quotient of the semistable locus.

Examples of GIT are Grassmannians, whose definition goes as follows. Take $G$ a complex (connected) reductive group. Let ${\cal O}$ denote the formal power series ring $\mathbf{C}[[z]]$ and ${\cal K}$ its fraction field $\mathbf{C}((z))$. The Grassmannian is defined by the group quotient 

\begin{equation}   
\text{Gr}_{_G}\ \overset{def.}{=}\ G_{_{\cal K}}/G_{_{\cal O}},   
\end{equation}
where $G_{_{\cal K}}$ and $G_{_{\cal O}}$ denote the groups of ${\cal K}$ and ${\cal O}$-valued points of $G$, respectively. More formally, Gr$_{_G}$ is defined as being the moduli space of pairs $({\cal P}, \varphi)$ of a $G$-bundle, ${\cal P}$, on the formal disk 

\begin{equation} 
D\ \overset{def.}{=}\ \text{Spec}({\cal O}), 
\end{equation}
and its trivialisation $\varphi$ over the punctured disk   

\begin{equation} 
D^{^*}\ \overset{def.}{=}\ \text{Spec}({\cal K}). 
\end{equation}    

Similarly, for the case of a complex representation, \textbf{N}, the moduli space is now parameterised by a triple

\begin{equation} 
{\cal R}\ \overset{def.}{=}\ ({\cal P}, \varphi, s),   
\end{equation}  
where $({\cal P}, \varphi)$ is in Gr$_{_G}$, and $s$ is a section of an associated vector bundle ${\cal P}_{_N}={\cal P}\times_{_G}\textbf{N}$, such that it is sent to a regular section of the trivial bundle under $\varphi$.

For a given triple ${\cal R}$, there is a corresponding projection

\begin{equation} 
\pi:\ {\cal R}\ \rightarrow \text{Gr}_{_G},     
\end{equation} 
where Gr$_{_G}$ is the affine Grassmannian, and $G$ is a complex reductive group. There is a natural commutative ring object, ${\cal A}$, in the derived $G_{_{\cal O}}$-equivariant constructible category on Gr$_{_G}$, $D_{_G}\left(\text{Gr}_{_G}\right)$, such that $H_{_{\bullet}}^{^{G_{_{\cal O}}}}\left({\cal R}\right)$ can be constructed from $H_{_{\bullet}}^{^{G_{_{\cal O}}}}\left(\text{Gr}_{_G},{\cal A}\right)$ equipped with a commutative ring structure.

In presence of a collection of such commutative ring objects, $\{{\cal A}_{_i}\}$, in $D_{_G}\left(\text{Gr}_{_G}\right)$, one can perform a \emph{gluing construction} by defining a new commutative ring object as 

\begin{equation}   
\iota _{_{\Delta}}\left(\ \boxtimes\ {\cal A}_{_i}\right),  
\end{equation}
where 

\begin{equation}  
\iota _{_{\Delta}}:\ \text{Gr}_{_G}\ \rightarrow\ \prod_{_i} \text{Gr}_{_G}  
\end{equation}
is the diagonal embedding. One such example are star-shaped quivers, \cite{Benini:2010uu}.

Its equivariant Borel-Moore homology can be defined by the $G_{_{\cal O}}$-equivariant Borel-Moore homology of   

\begin{equation}  
{\cal R}_{_{\le\lambda}}\ \equiv\ {\cal R}\ \cap\ \pi^{^{-1}}\left(\bar{\text{Gr}}_{_G}^{^{\lambda}}\right)     
\end{equation}    
in such a way that an embedding

\begin{equation}  
{\cal R}_{_{\mu}}\ \hookrightarrow\ {\cal R}_{_{\lambda}}   
\end{equation}  
will induce a map   

\begin{equation}  
H_{_{\bullet}}^{^{G_{_{\cal O}}}}\left({\cal R}_{_{\le\mu}}\right) \ \rightarrow\ H_{_{\bullet}}^{^{G_{_{\cal O}}}}\left({\cal R}_{_{\le\lambda}}\right) 
\end{equation}
for $\mu\le\lambda$.    

More generally, the euivariant Borel-Moore homology can be expressed as a graded sum 

\begin{equation}  
H_{_{\bullet}}^{^{G_{_{\cal O}}}}\left({\cal R}\right) \ \simeq\ \underset{\gamma}{\bigoplus}\  H_{_{\bullet}}^{^{G_{_{\cal O}}}}\left({\cal R}^{^{\gamma}}\right), 
\end{equation}
with ${\cal R}^{^{\gamma}}$ the connected component corresponding to $\gamma\ \in\ \pi_{_1}(G)$.

\section*{Coulomb and Higgs branches} 

The Coulomb branch is an affine algebraic variety whose coordinate ring is the equivariant Borel-Moore homology group, $H_{_{\bullet}}^{^{G_{_o}}}(\cal R)$ of a certain space ${\cal R}$, defining the variety of triples.

The Coulomb branch is defined as, \cite{Braverman:2016wma}, 

\begin{equation}    
{\cal M}_{_C}\ \overset{def.}{=}\ \text{Spec}\ {\cal A} (G, \mathbf{N})\ \equiv\ \text{Spec}\left(H_{_*}^{^{G_{_o}}}({\cal R}), *\right).     
\end{equation} 

Similarly, one can define the quantised Coulomb branch as follows  

\begin{equation}    
{\cal M}_{_C}\ \overset{def.}{=}\ \text{Spec}\ {\cal A}_{_{\hbar}} (G, \mathbf{N})\ \equiv\ \text{Spec}\left(H_{_*}^{^{G_{_o}\ \rtimes\mathbf{C}^{^{\times}}}}({\cal R}), *\right),     
\end{equation} 
which, in terms of quantum homology, can also be re-expressed as follows,

\begin{equation} 
{\cal M}_{_C}\ \overset{def.}{=}\ \text{Spec}\left(QH_{_G}^{^{\bullet}}\left(G;HF^{^{\bullet}}(X)\right)\right).     
\end{equation}

On th eother hand, the Higgs branch is defined as  the spectrum of the categorical quotient, namely, \cite{Braverman:2016wma},

\begin{equation} 
{\cal M}_{_H}\ \overset{def.}{=}\ \text{Spec}\left(QH^{^\bullet}\left(X//G\right)\right)\ \simeq\ QH^{^{\bullet}}_{_{LG}}(X).  
\end{equation}

\subsection{The monopole formula}

The monopole formula contains all the gauge-invariant chiral operators that acquire a non-zero expectation value\footnote{Their specific VEV depends on their respective dimensions and quantum numbers.} along the CB. It is particularly useful for investigating expected properties of Coulomb branches, and reads as follows   

\begin{equation}  
P_{_t}^{^{G_{_{\cal O}}}}({\cal R})\ =\ \sum_{_k}\ t^{^{-k}}\ \text{dim} \ H_{_k}^{^{G_{_{\cal O}}}}({\cal R}).    
\label{eq:MF}   
\end{equation}

As already remarked in \cite{Braverman:2016wma}, \eqref{eq:MF} is the same as the Hilbert series on the Coulomb branch of a 3D ${\cal N}=4$ supersymmetric quiver gauge theory associated with the pair $(G_{_{\mathbf{C}}}, \mathbf{N}\ \oplus\ \mathbf{N}^{^*})$.

\section{Back to the identity from moduli spaces}  \label{sec:3}

In this concluding section, we gather the key tools outlined in the present work, outlining some main features leading towards the relation between homological mirror symmetry and 3D mirror symmetry. 

3D mirror symmetry is defined as a duality exchanging Coulomb and Higgs branches, \cite{Intriligator:1996ex}. Making use of their algebro-geometric definition, briefly overviewed in section \ref{sec:32}, together with our previous work, \cite{Pasquarella:2023ntw}, we will be arguing that \emph{ordinary} 3D mirror symmetry, formulated as a 1-to-1 correspondence between a Coulomb and a Higgs branch, can be generalised to cases involving more theories (namely more Coulomb branches for a given Higgs branch). Specifically, we will be referring to the case of supersymmetric quiver gauge theories admitting a magnetic quiver description, and whose Higgs branch splits into two cones. This will ultimately provide further evidence to the claims made in \cite{Pasquarella:2023exd}.

As we shall see, the quantity that will enable us to probe such duality structure is the categorical interpretation of Hochschild cohomologies of block algebras, \ref{sec:fhss}.  

The present section is structured into three parts:

\begin{enumerate}

\item At first, we briefly overview the definition of block algebras, their associated Hoschschild cohomology, \cite{Ba}, and the relation with the Drinfeld center, the latter following from our preliminary discussion in section \ref{sec:334}.

\item  We then turn to Coulomb and Higgs branches arising from Hitchin system for class ${\cal S}$ theories, \cite{Neitzke:2014cja}. In particular, we highlight the fact that determining the Hitchin base, namely the Coulomb branch of the moduli space of vacua of class ${\cal S}$ theories, from the Higgs branch, is equivalent to specifying the identity in the embedding category, in complete analogy with the criterion outline din section \ref{sec:DCFMQS} and \ref{sec:334}.

\item Ultimately, we combine the tools outlined in the remainder of the present treatment, together with our previous work, \cite{Pasquarella:2023ntw}, explaining how in presence of algebraic varieties related to Moore-Tachikawa varieties without categorical duality, admit a generalised notion of 3D mirror symmetry thanks to the analogy with algebraic gauging in SymTFT constructions leading to intrinsically-non-invertible defects.

\end{enumerate}

\subsection{Hilbert series of Hochschild cohomology of block algebras}

In this first subsection, we will briefly overview a key result of \cite{Ba}, combining it with material outlined in section \ref{sec:DCFMQS}.

Let $p$ be a prime, $k$ an algebraically closed field of characteristic $p$, $G$ a finite group, and $B$ a block algebra of $kG$. A defect group of $B$ is a minimal subgroup $P$ of $G$ such that $B$ is isomorphic to a direct summand of $B\otimes_{_{kP}}B$ as a $B-B$-bimodule. The defect groups of $B$ form a $G$-conjugacy class of $p$-subgroups of $G$, and the defect of $B$ is the integer $d(B)$ such that $p^{^{d(B)}}$ is the order of the defect group of $B$.  

Donovan's conjecture predicts that there should only be finitely-many Morita equivalence classes of block algebras of Hochschild cohomology algebras of block algebras with a fixed defect $d$. For any integer $n\ \in\ \mathbb{N}$, dim $HH^{^n}(B)$ is bounded in terms of the defect groups of the block algebra $B$. 

The first result in \cite{Ba} adds to this the fact that there is a bound on the degrees of the generators and relations of the Hochschild cohomology $HH^{\bullet}(B)$, which only depends on the defect of $B$. In \cite{Ba}, this comes in three theorems, ultimately claiming that, the Hilbert series of the Hochschild cohomology of a block and its defect determine each other up to finitely many possibilities. In particular, 

\begin{equation} 
HH^{^{{\bullet}\ 0}}(B)\ \equiv\ \mathfrak{Z}(B),    
\end{equation}
where $\mathfrak{Z}(B)$ is the Drinfeld center.

\subsection{Hitchin systems and mirror quivers} 

We now turn to the more specific setup of interest to us, namely that of class ${\cal S}$ theories. In particular, we are interested in their moduli space of vacua, as well as their description in terms of 2D TQFTs, thanks to the AGT correspondence.

\subsection*{Hitchin systems}  

A Hitchin systems is an integrable system depending on the choice of a complex reductive group and a compact Riemann surface. It lies at the interface of algebraic geometry, Lie algebra theory, and integrable systems. It also plays a crucial role in the geometric Langlands program.  

The Hitchin equations consist in a system of PDEs on the Riemann surface, $C$, concerning a triple $(E, D, \varphi)$, \cite{Neitzke:2014cja}, where: 

\begin{enumerate} 

\item $E$ is a $G$-bundle on $C$.  

\item $D$ is a $G$-connection in $E$.  

\item $\varphi\ \in\ \Omega^{^1}\ (\text{End}\ E)$.  

\end{enumerate}

\subsection*{Spectral curves and Hitchin fibrations}  

The Higgs branch moduli space, ${\cal M}_{_H}$, is an example of one such complex integrable system. In particular, it is a fibration over a complex base space, ${\cal B}$, where the generic fiber is a compact complex torus. 

For gauge groups $G=SU(N)$ or $G=PSU(N)$, given a Higgs bundle $(E_{_h}, \phi)$, then the eigenvalues of $\phi$ in the standard representation of $G$ give an $N$-sheeted branch cover of $C$, \cite{Neitzke:2014cja},

\begin{equation}  
\Sigma\ \overset{def.}{=}\ \left\{\left(z\in C,\lambda\in T^{^*}_{_z}C\right):\text{det}(\phi(z)-\lambda)=0\right\}\ \subset\ T^{^*}C,      
\end{equation}    
which is the spectral curve corresponding to the Higgs bundle, $\left(E_{_h}, \phi\right)$.

The moduli space of harmonic bundles carries a natural family of complex structures $J_{_{\zeta}}$, parameterised by $\zeta\ \in\ \mathbf{CP}^1$. This follows form the fact that ${\cal M}$ carries a natural hyperk$\ddot{\text{a}}$hler metric, \cite{Neitzke:2014cja}. To see this, fix a $G$-bundle $E$. Then let ${\cal C}$ denote the space of pairs $(D,\varphi)$ without imposing the Hitchin equations. By definition, ${\cal C}$ is an infinite-dimensional affine space, with a natural hyperk$\ddot{\text{a}}$hler structure. Moreover, ${\cal C}$ is naturally acted on by the gauge group ${\cal G}$. This action preserves the hyperk$\ddot{\text{a}}$hler structure and has a moment map $\Vec{\mu}$. The Hitchin equations imply that all 3 components of $\Vec{\mu}$ must vanish. Denoting by ${\cal G}$ the smooth sections of Aut $E$, we therefore have

\begin{equation}  
{\cal M}\ \equiv\ \mu^{^{-1}}(0)/{\cal G},     
\end{equation}   
namely the hyperk$\ddot{\text{a}}$hler quotient 

\begin{equation}  
{\cal M}\ \equiv\ {\cal C}///{\cal G},   
\label{eq:hkq}   
\end{equation}  
implying ${\cal M}$ is in turn hyperk$\ddot{\text{a}}$hler. Every hyperk$\ddot{\text{a}}$hler manifold carries a canonical family of complex structures parametrised by $\zeta\  \in\ \mathbf{CP}^1$, in this case denoted by $J_{_{\zeta}}$.  

Let ${\cal B}$ denote the space of all $N$-sheeted branched covers $\Sigma\ \subset T^* C$ of $C$. In this case, ${\cal B}$ is a finite-dimensional complex vector space. Passing from the Higgs bundle to $\Sigma$ gives a projection known as the Hitchin fibration 

\begin{equation}  
{\cal M}_{_H}\ \rightarrow\ {\cal B}, 
\end{equation}
where ${\cal B}$ is the Hitchin base. 

As thoroughly explained in \cite{Neitzke:2014cja}, ${\cal B}$ is the Coulomb branch of the moduli space of vacua of the class ${\cal S}$ theory in question, conventionally denoted $X_{\mathfrak{g}}(C)$. At any nonsingular point in the Coulomb branch, there is an electromagnetic charge lattice 

\begin{equation} 
\Gamma\ \rightarrow\ {\cal B}_{_{reg}}\ =\ {\cal B}/{\cal B}_{sing}.  
\end{equation} 

At every point of base, $u\ \in\ {\cal B}$, there is a spectral curve, $\Sigma_{_u}$, thereby determining the local lattice system as the first cohomology, $H_{_1}(\Sigma_{_u},\mathbf{Z})$.

\subsection{Looking for the identity}  

In this concluding part of the section, we are now ready to gather all the ingredients outlined in the previous part of this work, to show how it relates to the formalism outlined in \cite{Pasquarella:2023ntw}. 

In doing so, we mostly need to recall the notion of the identity in Moore-Tachikawa varieties, and their generalisation for the case in which the target 2-category is a hyperk${\dddot{\text{a}}}$hler quotient

      \begin{equation}  
      T^{^*}\tilde G_{_{\mathbf{C}}}\ \overset{def.}{=}\ 
T^{^*}G_{_{\mathbf{C}}}^{^{\prime}}\ \otimes_{_{T^{^*}G_{_{\mathbb{C}}}}}\ T^{^*}G_{_{\mathbf{C}}}^{^{\prime\prime}}.
\label{eq:newidnow}
\end{equation} 

As explained in \cite{Pasquarella:2023ntw}, in such case, the identity is naturally removed by construction, and this clearly fits with taking the categorical quotient according to the GIT prescription, \eqref{eq:hkq}. From the work of Moore and Tachikawa, we also know that such categories are non-dualisable, and, consequently, not associated to a fully-extended TQFT.  

As explained in section \ref{sec:334}, full-extendibility can be recovered only upon specifying the identity element in the fully extended mother category from which the GIT quotient descends, \eqref{eq:prescr}. In particular, as suggested from \eqref{eq:newidnow}, the identity itself might be a composite object, as the one arising when performing double condensation of a SymTFT leading to non-invertible defects, \cite{Pasquarella:2023deo}.  

This partial gauging and gluing of different theories is precisely of the type occurring in the context of supersymmetric quiver gauge theories whose Higgs branch admits a magnetic quiver description, \cite{Cabrera:2019izd}, as well as 3D mirrors of Sicilian theories, \cite{Benini:2010uu}. We plan to report a more detailed treatment for these specific examples in due course, \cite{VP}.

\section{Conclusions and Outlook}

The present work is a followup to previous work by the same author, \cite{Pasquarella:2023exd,Pasquarella:2023ntw,Pasquarella:2023deo}. We take a more algebro-geometric approach in explaining the mathematical structure underlying supersymmetric quiver gauge theories, namely their dualities and their mutual relations in terms of higher-categories. As explained in depth, the crucial role is played by factorisation homology.  

As a biproduct of our treatment, we argued how it is possible to map the issue of gluing different class ${\cal S}$ theories separated by a non-invertible defect, \cite{Bashmakov:2022uek,Pasquarella:2023deo}, to that of defining the ground state degeneracy of a 2D topological order (TO) in presence of non-contractible loops. 

In summary, the present work was structured as follows: 

\begin{enumerate}

\item Section \ref{sec:DCFMQS} was devoted to explaining the role played by factorisation homology in describing the invariants of a given 2D TFT. In particular, this prepared the stage for the following sections, where to dealt with the applicability of this formalism to the specific case of class ${\cal S}$ theories.

\item In section \ref{sec:334}, we then turned to explaining the relation between categorical dualisability and full-extendibility, and how this nicely relates to the notion of Drinfeld centers and hoomological mirror symmetry.  In doing so, we focussed on the calculation of the Hilbert series as an algebraic variety of crucial interest in the study of categorical dualities for supersymmetric quiver gauge theories.

\item In section \ref{sec:32} we then introduced the notion of the Borel-Moore cohomology, as needed for defining the Grassmannian and the GIT quotient. After having recapitulated the essential tools required for performing GIT quotient constructions, we highlighted its importance for realising 3D mirror symmetry. 

\item The concluding section, namely \ref{sec:3}, shows how Coulomb and Higgs branches from class ${\cal S}$ Hitchin systems carry the same information as the gaugeable algebras leading to SymTFT constructions involving intrinsic-non-invertible symmetries, discussed in \cite{Pasquarella:2023deo}. We concluded outlining interesting open questions and future directions of investigation by the author, including further exploration of homological mirror symmetry for Sicilian, and star-shaped quiver gauge theories.

\end{enumerate}

\section*{Acknowledgements}

I wish to thank David Jordan for interesting conversations on related topics. Part of these findings were first presented remotely at the last conference organised by the Simons Collaboration It From Qubit at Perimeter, in July 2023.
This work is partially supported by the STFC Consolidated HEP theory grant ST/T000694/1 through DAMTP.

\appendix

\end{document}